\documentclass[fleqn]{elsart}

\usepackage[OT2,OT1]{fontenc}
\newcommand\cyr{%
\renewcommand\rmdefault{wncyr}%
\renewcommand\sfdefault{wncyss}%
\renewcommand\encodingdefault{OT2}%
\normalfont
\selectfont}
\DeclareTextFontCommand{\textcyr}{\cyr}
\usepackage{amsmath}
\usepackage{amsfonts}
\usepackage{amssymb}
\usepackage{graphicx}
\usepackage{bm}

\def\be{\begin{equation}}
\def\ee{\end{equation}}
\def\ba{\begin{eqnarray}}
\def\ea{\end{eqnarray}}
\def\bs{\begin{subequations}}
\def\es{\end{subequations}}

\def\rme{\text{e}}
\def\rmd{\text{d}}
\def\p{\partial}

\def\cL{{\cal L}}

\def\B{\Box}
\def\a{\alpha}

\def\s{\sigma}
\newcommand{\Eq}[1]{(\ref{#1})}

\begin{document}

\begin{frontmatter}

\title{Nonlocal instantons and solitons in string models}

\author{Gianluca Calcagni},
\ead{gianluca@gravity.psu.edu}
\address{Department of Physics and Astronomy, University of Sussex, Brighton BN1 9QH, United Kingdom}
\author{Giuseppe Nardelli},
\ead{nardelli@dmf.unicatt.it}
\address{Dipartimento di Matematica e Fisica, Universit\`a Cattolica,
via Musei 41, 25121 Brescia, Italia}
\address{INFN Gruppo Collegato di Trento, Universit\`a di Trento, 38100 Povo (Trento), Italia}

\begin{abstract}
We study a class of nonlocal systems which can be described by a local scalar field diffusing in an auxiliary radial dimension. As examples $p$-adic, open and boundary string field theory are considered on Minkowski, Friedmann--Robertson--Walker and Euclidean metric backgrounds. Starting from distribution-like initial field configurations which are constant almost everywhere, we construct exact and approximate nonlocal solutions. The Euclidean $p$-adic lump is interpreted as a solitonic brane, and the Euclidean kink of supersymmetric open string field theory as an instanton. Some relations between solutions of different string theories are highlighted also thanks to a reformulation of nonlocal systems as fixed points in a renormalization group flow.
\end{abstract}



\end{frontmatter}


\section{Introduction and main results}

The treatment of nonlocal systems, which contain an infinite number of derivatives, is of some interest for string field theory (bosonic, supersymmetric and $p$-adic). In fact, time-dependent solutions of the tachyon mode rolling from the unstable (perturbative) vacuum to the stable (non-perturbative) vacuum have been sought for long time. Boundary string field theory (BSFT) possesses such solutions. On the other hand, $p$-adic string in a suitable limit reproduces BSFT, so that it is natural to expect instanton-like analytic solutions (or, more generally, non-perturbative topologically stable solutions) in the $p$-adic model. In these scenarios the main equation of motion of the tachyon mode is always
\be\label{1}
{\cal K}\rme^{-2r_*\B}\phi(x)=U'[\phi(x)]\,,
\ee
where ${\cal K}$ is the kinetic operator determined by the nonlocal model, $r_*$ is a constant and $\B$ is the d'Alembertian operator in $D$ dimensions with metric signature $({-}{+}\cdots{+})$. Acting on a scalar it reads $\B=(-g)^{-1/2}\p_\mu(\sqrt{-g}\,\p^\mu)$, where $g$ is the determinant of the metric and $\mu=0,1,\dots,D-1$.

This class of models can be conveniently recast as a two-field \emph{localized} system living in $1+D$ dimensions, characterized by the spacetime coordinates $x^\mu$ and an auxiliary radial direction $r$ replacing the `physical' parameter $r_*$ \cite{cuta3}. The resulting laws of motion are the heat (or diffusion) equation
\be
(\B-\gamma\p_r)\Phi(r,x)=0\,,\label{difeq}
\ee
where $\gamma$ is a constants, and the proper equation of motion
\be
{\cal K}\Phi(r_*-2\gamma r_*,x) = U'[\Phi(r_*,x)]\,,\label{eom}
\ee
which is valid only at $r=r_*$ and where $\Phi(r_*,x)=\phi(x)$.\footnote{In the notation of \cite{cuta3}, the auxiliary field $\chi$, which we do not consider here for simplicity, is defined as $\chi(r,x)={\cal K}\Phi(r,x)$.} Applications of this formalism were given for the construction of tachyon solutions in open string field theory (OSFT; the bosonic, and sometimes also the supersymmetric, version is called also cubic SFT) \cite{roll} and cosmological toy models \cite{cuta2}. The main idea is to evolve the system, via Eq.~\Eq{difeq}, from the initial field configuration $\Phi(0,x)$ , up to the physical configuration $\Phi(r_*,x)$. Nonlocal operators $\rme^\B$ act as translations on the variable $r$. 

If $\Phi(0,x)$ is assumed to be a \emph{nontrivial} solution of the \emph{local} system ($r=0$ everywhere), it is possible to construct $\Phi(r_*,x)$ analytically, as exact local solutions are generally known for a given potential. However, the resulting nonlocal solutions for the models considered so far are approximate at the level either of the scalar equation of motion \Eq{eom} \cite{roll,FGN} or of the Friedmann equation if the metric is cosmological \cite{cuta2}. Excluding an a priori failure of the method, which has been supported by encouraging results (see also \cite{vla05}), the reason may be an unsuitable choice of the initial field configuration at $r=0$.

An alternative simple choice is that of \emph{almost everywhere} constant local solutions, with possible singularities or discontinuities at a finite number of points. For instance, the $p$-adic string \cite{FO,FW,BFOW} (see also \cite{vla05,VV,vla07}) admits $\Phi=0$ as constant solution, so that $\Phi(0,x)=\delta(x)$ is a local solution everywhere except at the origin. By diffusion along $r$, we shall see that the final configuration $\Phi(r_*,x)$ is a smooth Gaussian lump which solves the original, nonlocal system.

The $p$-adic local equation of motion has also $\Phi=\pm 1$ as solutions for odd $p$, and $\Phi(0,x)={\rm sgn}(x)$ may be used as initial configuration. Although the resulting kink cannot be a solution of the $p$-adic equation, a similar construction is successful in the case of supersymmetric OSFT (see the reviews \cite{ohm01,ABGKM,sen04} and references therein). On the other hand, the lump construction does not hold in OSFT for reasons to be clear later. However, the same lump solves the effective Witten--Shatashvili action (up to second derivatives of the field) in boundary string field theory (BSFT) \cite{wit1,wit2,sha1,sha2}, which can then be regarded as a local diffusing system. This datum adds to the findings of \cite{GS}, where a relation was noted between the $p$-adic string in the limit $p\to 1$ and Witten--Shatashvili action. We will also show how to the $p$-adic system can be regarded as the limit of non-propagating (in a sense later to be clarified) OSFT.

The aim of this Letter is twofold: to incorporate the $p$-adic string in the discussion of diffusing models and find solutions of the $p$-adic string as well as of OSFT with the above-mentioned new type of initial conditions. As a byproduct, interpreting the diffusion equation as a renormalization group flow, we shall find a formal connection between $p$-adic and open string field theories.

We carry out these investigations beginning with covariant equations; the structure of the diffusing system and ansatz solutions soon require to specialize to particular backgrounds. Lumps will be studied on Minkowski, Friedmann--Robertson--Walker (FRW) and Euclidean backgrounds, the latter being also the natural setting for kink solutions.


\section{Lump solutions}\label{lumps}

We look for solutions of the form
\be\label{sol1}
\Phi(r,x)=\Phi_0\,r^{-c/2} \exp\left[-\frac{\gamma \a T(x)}{4r}\right]\,,
\ee
where $\Phi_0$, $c$ and $\a$ are constants and $T(x)$ is a function of the coordinates. On a Minkowski background,
\be\label{qua}
T(x)= x^2_{\bot}\equiv x_A x^A\,,
\ee
where the index $A$ runs over a subset $S\subseteq \{0,1,\dots,D-1\}$. If the cardinality ${\rm card}(S)=c$ and $\a\gamma/r>0$, the Euclidean continuation of $\Phi$ is a product of Gaussians, which tends to the Dirac distribution $\delta^{(c)}$ in the limit $r\to 0$.

By virtue of the heat equation \Eq{difeq}, the d'Alembertian $\B$ can be replaced by a derivative in $r$, thus allowing for a covariant treatment independent of the choice of the metric. Namely, one can consider a general function $T(x)$ such that Eq.~\Eq{sol1} satisfies the diffusion equation, which is\footnote{In \cite{cuta3} there is an extra linear term in Eq.~\Eq{difeq} as the constant $\Phi_0$ is considered to depend on $r$. However, for all purposes $\Phi_0$ can be treated as $r$-independent, $\Phi_0=\Phi_0(r_*)$, and the extra term can be neglected without loss of generality.}
\be\label{diffi}
\B\Phi(r,x)=-\frac{\gamma}{2r}\Phi(r,x)\left[u_0+\ln\Phi^2(r,x)\right]\,,
\ee
where $u_0\equiv c-\ln(\Phi_0^2 r^{-c})$. This translates into an equation for $T$ and the metric $g_{\mu\nu}$ of the form $A(T,g)+rB(T,g)=0$. As $r$ is a free parameter in the localized system, $A(T,g)$ and $B(T,g)$ must vanish separately, namely,
\ba
&&\a\B T-2c=0\,,\label{cons}\\
&&\a\p_\mu T\p^\mu T-4T=0\,.
\ea
These equations are verified only when $T$ is quadratic in the coordinates, a result stemming directly from dimensional counting of the variables in the diffusion equation. Apart from Minkowski and Euclidean backgrounds (Eq.~\Eq{qua} with $\a=1$, ${\rm card}(S)=c$ \cite{BFOW}) another possibility is FRW ($\B=-\p_t^2-(D-1)H\p_t$), where $\a T=-t^2$ and Eq.~\Eq{cons} fixes the Hubble parameter to be $H=H_0/t$, $H_0=(c-1)/(D-1)$. As expected, the Minkowski case is recovered when $c=1$, where Eq.~\Eq{sol1} is the usual heat kernel. If $c\ne 1$, Eq.~\Eq{sol1} solves the heat equation in the above FRW background, and for $r\to 0$ it tends to the delta function (with the appropriate metric factor in the measure), so that Eq.~\Eq{sol1} is  the heat kernel in the above FRW background.
 

\subsection{$p$-adic string}

The $p$-adic Lagrangian is \cite{BFOW}
\be
\cL = -\frac12\phi p^{-\B/2}\phi+\frac{\phi^{p+1}}{p+1}\,,
\ee
plus a possible constant term. The equation of motion can be written as
\be\label{eo}
\rme^{-\ln p\B/2}\phi=\phi^p.
\ee
In $1+D$ notation, the corresponding localized system has 
\be\label{pad}
r_*=r_p\equiv(\ln p)/4\,,\qquad {\cal K}=1\,,\qquad U'=\Phi^p\,,
\ee
so that $\rme^{-2r_p\B}\Phi=\Phi^p$. 
The system always admits the constant solutions
 $\Phi=0,\pm 1$ ($-1$ only if $p$ is odd). We shall consider the solution $\Phi=0$ (almost everywhere) as the initial field configuration to be evolved to Eq.~\Eq{sol1} through the diffusion equation. 

The $p$-adic equation of motion becomes the algebraic equation \Eq{eom}. Powers of $\Phi$ translate under the effect of nonlocal operators, as
\be
\Phi(r,x)^p= (p r^{p-1})^{-c/2} \Phi_0^{p-1}\Phi(r/p,x)\,.
\ee
Then one gets $\gamma =(p-1)/(2p)$ and fixing the scale $r=r_p$ the $p$-adic solution is
\be\label{padicsol}
\Phi(r_*,x)= p^{c/2(p-1)}\exp\left[-\frac{(p-1)\a T(x)}{2p\ln p}\right]\,,
\ee
in agreement with \cite{BFOW} when $T$ is given by Eq.~\Eq{qua}, $\a=1$ and ${\rm card}(S)=c$. The static version of this solution ($\{0\}\not\subset S$) vanishes asymptotically for $p>0$, and is interpreted as a solitonic $p$-adic brane of codimension $c$ \cite{GhS}.

A rapid overview of the energy-momentum tensor \cite{cuta3} in a given cosmological scenario shows that this solutions of the scalar equation of motion does not obey a standard Friedmann equation $H\propto \rho^q$, where the Hubble parameter $H$ and the field energy density $\rho$ evolve together ($q$ positive). This is the case also when one allows for other forms of the Hubble parameter, for instance by multiplying the right-hand side of Eq.~\Eq{sol1} times a power of $t$. The construction of an ad-hoc nonperturbative and nonlocal gravitational sector is beyond the scope of this Letter and, so far, of the above formalism. It is possible that the na\"ive local Friedmann equation does not correctly describe nonperturbative gravity and, until a realistic proposal for the latter is put forward, one should content oneself either of numerical solutions \cite{jou07} or to pursue the more modest goal of finding exact solutions of the scalar equation of motion \Eq{eom} for a given background. This will be our attitude on what follows.

It is easy to see that solutions with asymptotics $\Phi(r,x\to +\infty)=\Phi(r,x\to -\infty)\neq 0$ cannot be written down as Eq.~\Eq{padicsol} plus a constant and are better to be found numerically \cite{jou07,MZ,vol03,jou2}. For cosmological applications of the $p$-adic scalar see \cite{jou07,BBC}.


\subsection{Supersymmetric OSFT}\label{als}

We now turn our attention to Eq.~\Eq{1} with kinetic operator (Minkowski background)
\be\label{2}
{\cal K}=\B-m^2\,.
\ee
In the case of nonchiral bilocal supersymmetric OSFT \cite{PTY,AKBM}, $m^2=-1/2$, $r_*\equiv \ln(3^{3/2}/4)\approx 0.26$, and
\be\label{eox}
U'=\frac{\rme^{4r_*}}{9} \Phi\,\rme^{2r_*\B}\Phi^2\,
\ee
at level $(1/2,1)$ (for details see \cite{AJK}). For convenience we define
\be
\s\equiv\frac{\rme^{4r_*}}{9} \Phi_0^2\,,
\ee
and
\be
\Delta\equiv \left|\frac{{\cal K}\Phi_{\rm app}-U'(\Phi_{\rm app})}{{\cal K}\Phi_{\rm app}+U'(\Phi_{\rm app})}\right|\,,
\ee
where $\Phi_{\rm app}$ is an approximate solution. 

One of the advantages of the localization achieved through the heat equation is the possibility to construct solutions which, if they are not global, can be studied locally. Here we give two such examples; others can be found in \cite{roll,cuta2}.

The first one is given by Eq.~\Eq{sol1} with $\a T=x^2$ (one-dimensional spatial lump). A complete lump solution would describe the rolling of the tachyon from the local unstable vacuum down to a local minimum and back. However, Eq.~\Eq{sol1} is not an exact solution of this model unless $T={\rm const}$. At most, one can tune $\Phi_0$ and $\gamma$ in order to get an approximate solution in a neighborhood of $x=0$, the values of the parameters depending on the accuracy goal within a given neighborhood. For instance, when $\s\approx 1.20$ and $\gamma\approx -1.34$, $\Delta<0.5\%$ for $|x|<1$.


\subsection{Boundary string field theory as a diffusing system}


At this point it is instructive to ask what potential would admit Eq.~\Eq{sol1} as a solution of the equation of motion of the localized system with 
\be
{\cal K}=\B.
\ee
This problem is better recast as follows. Taking Eq.~\Eq{sol1} as starting point, the diffusion equation for $\Phi$ is Eq.~\Eq{diffi}. Fixing $r=r_*$, it becomes the second-order equation for the field $\phi(x)$,
\be\label{diffi2}
\B\phi=-\frac{\gamma}{2r}\phi\left(u_0+\ln\phi^2\right)\,,
\ee
which comes from the \emph{local} Lagrangian density
\be\label{lagf}
\cL_\phi=4\p_\mu\phi\p^\mu\phi+\frac{2\gamma}{r_*}\phi^2\left(1-u_0-\ln\phi^2\right)\,.
\ee
If
\be
\a = \frac{2r_*}{\gamma}\,,
\ee
Eq.~\Eq{lagf} can be recast as
\be\label{lagT}
\cL_T=\rme^{-T}\left[\p_\mu T\p^\mu T+\frac{2\gamma}{r_*}(1-c+T)\right]\,.
\ee
Provided
\be
c=0\,,\qquad \gamma=r_*/2\,,
\ee
this is nothing but the bosonic tree-level effective action of the BSFT tachyon, when all other particle fields are set to zero and up to two derivatives \cite{GS,KMM1,CFGNO}.


\section{Kink solutions}

Given a potential with several local minima, one can study kink-type Euclidean solutions interpolating between two different vacua. The tunneling probability is then related with the effective Euclidean action evaluated on the solution.

For an even potential with local minima at $\Phi=\pm 1$, one can start from the field configuration $\Phi(0,x)={\rm sgn}(x)$, which is the limit $r\to 0$ of the error function
\be\label{ers}
\Phi(r,x)={\rm erf}\left(\pm\sqrt{\frac{\gamma}{4r}}\,x\right)\,,
\ee
where (\cite{GR}, formul\ae\ 8.250.1 and 8.253.1)
\ba
{\rm erf}(z)&\equiv& \frac{2}{\sqrt{\pi}} \int_0^z\rmd\s \rme^{-\s^2}\label{erf}\\
      &=& \frac{2}{\sqrt{\pi}} \sum_{k=0}^{\infty} (-1)^k \frac{z^{2k+1}}{k!(2k+1)}\,.
\ea
For $z\to\pm\infty$ its asymptotic expansion is (\cite{GR}, formula 8.254)
\be\label{asi}
{\rm erf}(z) \stackrel{z\to \pm\infty}{\sim} \pm 1-\frac{\rme^{-z^2}}{\sqrt{\pi}\,z}\sum_{k=0}^M (-1)^k\frac{(2k-1)!!}{(2z^2)^k}\,.
\ee
Equation \Eq{ers} obeys the diffusion equation \Eq{difeq}.

For the $p$-adic string, approximate kink-type solutions, such that $\Phi(r,t\to +\infty)\neq\Phi(r,t\to -\infty)$, have been found in \cite{vla05}. From Eq.~\Eq{asi} it is clear that the error function cannot be an exact solution of the $p$-adic equation \Eq{eo}. In fact, its left-hand side is $\sim x$ for small $x$, while the right-hand side $\sim x^p$. One might look for a solution locally at large $x$ but we prefer to turn our attention to the OSFT case.

In this case, the presence of the d'Alembert operator in the equation of motion, Eqs.~\Eq{1} and \Eq{2}, makes possible a matching of powers of $x$ between the sides of the equation of motion and the construction of an approximate solution. As $\B\Phi\to 0$ in the limit $x\to\pm\infty$, by virtue of Eq.~\Eq{asi} $\s$ is fixed by the asymptotes at infinity to be $\s=-m^2=1/2$. 

As powers of the error function do not obey a diffusion equation, from now on we concentrate on the approximated susy potential \cite{AJK}
\be\label{apex}
U'\approx \frac{\rme^{4r_*}}{9} \Phi^3\,.
\ee 
Expanding $\Phi$ and its equation of motion near the origin, one finds
\be
\gamma=\frac{r_*}{1+2r_*}\approx 0.17\,.
\ee
With these values of $\s$ and $\gamma$, the error $\Delta$ is maximal ($\Delta\approx 1.5\%$) at $x\sim \pm 3$, while $\Delta<0.1\%$ and decays to zero for $|x|\gtrsim 5$. Therefore this is an approximate global solution (on the other hand, the lump approximate solution in Section \ref{als} is acceptable only in a limited space interval). Numerical kinks solving the susy OSFT equation can be found in \cite{jou07,vol03}.


\section{Effective string models as diffusing systems}

The diffusing system in $1+D$ dimensions can be reinterpreted in the language of the renormalization group (RG) flow. First of all, $r$ is regarded as the scale of an RG flow, $r=0$ being the point at the infrared (IR) where the theory is under control (local regime) and $r=r_*$ is the physical ultra-violet (UV) scale at which observations take place and nonlocal effects are measurable. The operator $\rme^{s\B}$ evolves the field $\Phi(r,t)$ from some point $r$ to a point $r+\gamma s$ along the RG flow. (In the above example of lump solutions, the initial pointwise particle has been `resolved' as a spread object in the UV.) A truncation of this operator at finite order is equivalent to a perturbative expansion which diverges in the ultra-violet. The resummation with the diffusion equation method solves any spurious problems of UV convergence.

Further insight is gained by looking at the propagator $G_r$ of the free field $\Phi$, Eq.~\Eq{1} with $U=0$ and a typical kinetic function ${\cal K}=\B$. In the infra-red local regime and in Euclidean momentum space ($\B\to -k_E^2<0$), it is simply $G_0(k_E)=-1/k_E^2$. As the system evolves along the flow, the propagator acquires corrections which are nonperturbatively resummed into
\be
G_{r_*}(k_E)= G_0(k_E)\,\rme^{-2r_*k_E^2}\,.
\ee
If $r_*>0$, a characteristic UV cutoff $k_c\equiv (2r_*)^{-1/2}$ naturally emerges from the flow, and modes with momentum $k_E\gg k_c$ are exponentially suppressed. These modes were responsible of the UV perturbative divergences, as is clear by expanding the propagator $G(r_*)$ in powers of $k_E/k_c$.\footnote{A general class of nonlocal quantum field theories (NQFT), of which Eq.~$\Eq{1}$ is a special example, was axiomatized and studied in detail by Efimov \cite{efi77}. In the terminology of NQFT, the presence of the coherence scale $k_c$ guarantees a well-defined nonperturbative physics and the system is actually super-renormalizable. Although this view of the facts has become obsolete in string theory (the fundamental object to be quantized is a field of strings; the effective theory given by Eq.~\Eq{1} is purely classical), it retains an appreciable degree of instructiveness, particularly in early-universe cosmological contexts where one is interested in the power spectr
 um of the scalar field.}

As an immediate application of these remarks, it is worth noticing that the $p$-adic and OSFT systems can be regarded as different fixed points in the flow along the $r$ coordinate. Under the coordinate transformation $x^\mu\to \sqrt{r_*/r}\,x^\mu$, one can write the OSFT equation of motion (approximately, for the susy string) as\footnote{For numerical purposes, the same type of equation was considered in \cite{vol03,AJK}.}
\be
\left(\frac{r}{r_*}\B-m^2\right)\rme^{-2r\B}\Phi=\s\Phi^p\,,
\ee
The limit $r\to r_*\approx 0.26$ corresponds to the OSFT tachyon equation of motion in the old coordinates. However, in the singular limit of no propagation, $r_*\to\infty$, one formally gets the $p$-adic equation (for a suitable field normalization) if the fixed point in the $r$-flow is $r=r_p$. 

This closes the set of correspondences between $p$-adic string theory, BSFT and OSFT considered here and in \cite{roll}. To summarize: (A) The $p$-adic string can be seen as the limit of a non-propagating OSFT tachyon, where the UV cut-off $k_c\to 0$. This is true on any nontrivial solution. (B) The $p$-adic string reproduces the BSFT action in the limit $p\to 1$ \cite{GS} for any tachyon profile $T$. On the other hand, the effective action whose equation of motion is the diffusion equation is precisely that of BSFT up to two derivatives and with a tachyon profile $T\sim x^2$. (C) The partition function (effective action) of BSFT evaluated on the tachyon profile $T\sim \rme^{x}$ and with propagator ambiguity $r_*>0$ is related  to the (approximate) oscillating OSFT solution as described in \cite{roll}. These results are readily obtained by regarding all the systems as diffusing. The interpretation of such relations will require further investigation, as at this point it is not clear whether their valence is physical or only technical.


\section{Discussion}

We have studied  nonlocal systems which can be localized via the diffusion equation, as described in \cite{cuta3}. The main points are:
\begin{itemize}
\item Contrary to what done in previous papers \cite{roll,cuta2}, we have assumed the initial field configuration to be constant almost everywhere, leading to lump- and kink-type solutions. 
\item The solitonic solution in $p$-adic string theory, which was already known in the literature, has been here  rederived as an almost trivial application of the formalism. 
\item The same type of solution is of interest for the Witten--Shatashvili effective action of boundary SFT; the second-order equation of motion of the BSFT tachyon does coincide with the heat equation frozen at a particular value of the variable $r$.
\item We found an approximate analytic instantonic solution interpolating between two degenerate minima of the supersymmetric OSFT tachyon potential.
\item Regarding $p$-adic, OSFT and BSFT tachyon models as diffusing/RG-flow systems, one readily obtains relations between the corresponding solutions.
\end{itemize}

The comparison between approximate analytic solutions and numerical solutions with same boundary conditions can help to understand capabilities and limitations of the diffusion method. The case of kink-type OSFT profiles is of particular interest, as they would describe a nonperturbative transition between inequivalent vacua of the theory. Moreover, since $p$-adic models have been used as simplified playgrounds for the study of nonlocality in inflationary models, the search for background solutions could be of practical importance for alternative cosmological scenarios.


\ack

G.C. thanks the Department of Mathematics and Physics of Universit\`a Cattolica for the kind hospitality during the completion of this Letter. The work of G.C.\ is supported by a Marie Curie Intra-European Fellowship under contract MEIF-CT-2006-024523. G.N.\ is partly supported by INFN of Italy.
 


\begin{thebibliography}{100}
\bibitem{cuta3} G. Calcagni, M. Montobbio, G. Nardelli, Phys. Lett. B 662 (2008) 285, arXiv: 0712.2237 [hep-th].
\bibitem{roll}  G. Calcagni, G. Nardelli, arXiv: 0708.0366 [hep-th].
\bibitem{cuta2} G. Calcagni, M. Montobbio, G. Nardelli, Phys. Rev. D 76 (2007) 126001, arXiv: 0705.3043 [hep-th].
\bibitem{FGN}   V. Forini, G. Grignani, G. Nardelli, JHEP 0503 (2005) 079, hep-th/0502151.
\bibitem{vla05} V.S. Vladimirov, math-ph/0507018.
\bibitem{FO}    P.G.O. Freund, M. Olson, Phys. Lett. B 199 (1987) 186.
\bibitem{FW}    P.G.O. Freund, E. Witten, Phys. Lett B 199 (1987) 191.
\bibitem{BFOW}  L. Brekke, P.G.O. Freund, M. Olson, E. Witten, Nucl. Phys. B 302 (1988) 365.
\bibitem{VV}    V.S. Vladimirov, Ya.I. Volovich, Theor. Math. Phys. 138 (2004) 297, math-ph/0306018.
\bibitem{vla07} V.~Vladimirov, Theor.\ Math.\ Phys.\  149 (2006) 1604, arXiv: 0705.4600 [math-ph].
\bibitem{ohm01} K. Ohmori, hep-th/0102085.
\bibitem{ABGKM} I.Ya. Aref'eva, D.M. Belov, A.A. Giryavets, A.S. Koshelev, P.B. Medvedev, hep-th/0111208.
\bibitem{sen04} A. Sen, Int. J. Mod. Phys. A 20 (2005) 5513, hep-th/0410103.
\bibitem{wit1}  E. Witten, Phys. Rev. D 46 (1992) 5467, hep-th/9208027.
\bibitem{wit2}  E. Witten, Phys. Rev. D 47 (1993) 3405, hep-th/9210065.
\bibitem{sha1}  S. Shatashvili, Phys. Lett. B 311 (1993) 83, hep-th/9303143.
\bibitem{sha2}  S. Shatashvili, Alg. Anal. 6 (1994) 215, hep-th/9311177.
\bibitem{GS}    A.A. Gerasimov, S.L. Shatashvili, JHEP 0010 (2000) 034, hep-th/0009103.
\bibitem{GhS}   D.~Ghoshal, A.~Sen, Nucl.\ Phys.\  B 584 (2000) 300, hep-th/0003278.
\bibitem{jou07} L.V. Joukovskaya, Phys.\ Rev.\  D 76 (2007) 105007, arXiv: 0707.1545 [hep-th].
\bibitem{MZ}    N.~Moeller, B.~Zwiebach, JHEP 0210 (2002) 034, hep-th/0207107.
\bibitem{vol03} Ya.I. Volovich, J. Phys. A 36 (2003) 8685, math-ph/0301028.
\bibitem{jou2}  L.V.~Joukovskaya, Theor.\ Math.\ Phys.\ 146 (2006) 335, arXiv: 0708.0642 [math-ph].
\bibitem{BBC}   N. Barnaby, T. Biswas, J.M. Cline, JHEP 0704 (2007) 056, hep-th/0612230.
\bibitem{PTY}   C.R. Preitschopf, C.B. Thorn, S.A. Yost, Nucl. Phys. B 337 (1990) 363.
\bibitem{AKBM}  I.Ya. Aref'eva, A.S. Koshelev, D.M. Belov, P.B. Medvedev, Nucl. Phys. B 638 (2002) 3, hep-th/0011117.
\bibitem{AJK}   I.Ya. Aref'eva, L.V. Joukovskaya, A.S. Koshelev, JHEP 0309 (2003) 012, hep-th/0301137.
\bibitem{KMM1}  D.~Kutasov, M.~Mari\~no, G.W.~Moore, JHEP 0010 (2000) 045, hep-th/0009148.
\bibitem{CFGNO} E. Coletti, V. Forini, G. Grignani, G. Nardelli, M. Orselli, JHEP 0403 (2004) 030, hep-th/0402167.
\bibitem{GR}    I.S. Gradshteyn, I.M. Ryzhik, Table of Integrals, Series, and Products, Academic Press, London, 2000.
\bibitem{efi77} G.V. Efimov, \textcyr{Nelokal\cyrsftsn{}nye vzaimode\U{i}stviya kvantovannykh pole\U{i}} (Nonlocal interactions of quantized fields), Nauka, Moscow, 1977.
\end{thebibliography}
\end{document}